\documentclass[superscriptaddress,showpacs,preprintnumbers,preprint,amsmath,amssymb]{revtex4}

\usepackage{graphicx}
\usepackage{epstopdf}
\usepackage{bm}
\usepackage{color}

\newcommand{\Sergey}[1]{\textcolor{blue}{{#1}}}

\newcommand{\comment}[1]{}
\def\be{\begin{equation}}
\def\ee{\end{equation}}
\def\bea{\begin{eqnarray}}
\def\eea{\end{eqnarray}}
\def\noi{\noindent}

\begin{document}

\preprint{APS/123-QED}

\title{Photon echo transients from an inhomogeneous ensemble\\ of semiconductor quantum dots}

\author{S.~V.~Poltavtsev}
    \email{svp@bk.ru}
	\affiliation{Experimentelle Physik 2, Technische Universit\"at Dortmund, 44221 Dortmund, Germany}
    \affiliation{Spin Optics Laboratory, St.~Petersburg State University, 198504 St.~Petersburg, Russia}
    \author{M.~Salewski}
    \affiliation{Experimentelle Physik 2, Technische Universit\"at Dortmund, 44221 Dortmund, Germany}
\author{Yu.~V.~Kapitonov}
    \affiliation{St.~Petersburg State University, 199034 St.~Petersburg, Russia}
\author{I.~A.~Yugova}
    \affiliation{Spin Optics Laboratory, St.~Petersburg State University, 198504 St.~Petersburg, Russia}
\author{I.~A.~Akimov}
    \affiliation{Experimentelle Physik 2, Technische Universit\"at Dortmund, 44221 Dortmund, Germany}
    \affiliation{Ioffe Physical-Technical Institute, Russian Academy of Sciences, 194021 St.~Petersburg, Russia}
\author{C.~Schneider}
\author{M.~Kamp}
\author{S.~H\"ofling}
    \affiliation{Technische Physik, Universit\"at W\"urzburg, 97074 W\"urzburg, Germany}
\author{D.~R.~Yakovlev}
    \affiliation{Experimentelle Physik 2, Technische Universit\"at Dortmund, 44221 Dortmund, Germany}
    \affiliation{Ioffe Physical-Technical Institute, Russian Academy of Sciences, 194021 St.~Petersburg, Russia}
\author{A.~V.~Kavokin}
    \affiliation{Spin Optics Laboratory, St.~Petersburg State University, 198504 St.~Petersburg, Russia}
    \affiliation{School of Physics and Astronomy, University of Southampton, SO17 1 BJ, Southampton, United Kingdom}
    \affiliation{CNR-SPIN, Viale del Politecnico 1, I-00133 Rome, Italy}
\author{M.~Bayer}
    \affiliation{Experimentelle Physik 2, Technische Universit\"at Dortmund, 44221 Dortmund, Germany}
    \affiliation{Ioffe Physical-Technical Institute, Russian Academy of Sciences, 194021 St.~Petersburg, Russia}

\date{\today}

\begin{abstract}
An ensemble of quantum dot excitons may be used for coherent
information manipulation. Due to the ensemble inhomogeneity any
optical information retrieval occurs in form of a photon echo. We
show that the inhomogeneity can lead to a significant deviation from
the conventional echo timing sequence. Variation of the area of the
initial rotation pulse, which generates excitons in a dot
sub-ensemble only, reveals this complex picture of photon echo
formation. We observe a retarded echo for $\pi/2$ pulses, while for
$3\pi/2$ the echo is advanced in time as evidenced through
monitoring the Rabi oscillations in the time-resolved photon echo
amplitude from (In,Ga)As/GaAs self-assembled quantum dot structures and
confirmed by detailed calculations.
\end{abstract}



\maketitle


Manipulation of electronic states by laser pulses requires precise
timing of the pulse sequence. A particular example is ultrafast
coherent control of exciton complexes confined in semiconductor
nanostructures \cite{Yamamoto13} which may be used for manipulating
quantum states \cite{GreilichNatPhys,Biolatti,Li} or even storing
quantum information \cite{Kroutvar,LangerNature}. For an
inhomogeneous ensemble of two-level systems (TLS) such as quantum
dots (QDs) the information retrieval induces echoes for the
recovered signal~\cite{LangbeinPRL2001,Moody2013,TaharaPRB2014}. To
be of practical use, the echo formation time needs to be determined
with high accuracy.

The scenario is well established when a weakly inhomogeneous
ensemble of TLS is resonantly addressed by ultrashort laser pulses
whose spectrum covers the distribution of optical transitions in the
ensemble. In this conventional picture of a photon echo (PE), a
first excitation pulse creates a macroscopic polarization which
subsequently dephases due to inhomogeneity. A second subsequent
optical rephasing pulse retrieves the macroscopic polarization
\cite{Shen} with subsequent emission of a PE pulse delayed by a time
exactly equal to the interval between the rephasing and excitation
pulse, $\tau_{12}$. Typically the temporal profile of the PE pulse
has a Gaussian shape reflecting the inhomogeneous frequency
distribution of the excited ensemble.

In semiconductors this picture may be complicated by several
features: (1) Due to the many-body interactions a description by a
distribution of independent TLS is as a rule not appropriate, but
complex coherent transients are recorded \cite{Chemla2001,ShahBook}.
However, localization of excitons strongly reduces exciton-exciton
interactions \cite{NollPRL}. Therefore, QDs with particularly strong
three-dimensional localization can be often reasonably well
considered as TLS with long coherence times. Indeed, Rabi
oscillations of exciton complexes were demonstrated in QDs using various
techniques \cite{Zrenner2002,GammonPRL,RamsayPRL,Gamouras}. (2) In
addition, fluctuations in QD size, shape and composition lead to a
considerable inhomogeneous broadening of the optical transitions in
QDs. Simultaneously excitation with ultrashort pulses having a
broader spectrum than the inhomogeneity in many cases is prohibited
because of the demand of addressing a specific optical transition.
The relatively narrow energy spacing of different electronic
excitations limits the spectral width that the laser pulses may
have, to avoid excitation of multiple transitions.

Therefore a careful consideration of the optical frequency detuning
with respect to the TLS optical resonance is mandatory when
considering the PE transient formation, even in the limit of small
excitation levels. Generally, it is expected that for strongly
inhomogeneous TLS ensembles the PE temporal profile may undergo
significant distortion compared to the standard
$\tau_{12}-2\tau_{12}$ sequence, showing either retarded or advanced
echo pulses \cite{AllenEberly}. This is due to the non-negligible
inhomogeneity-induced dephasing of oscillators during the optical
excitation pulse, so that the finite pulse duration should be taken
into account. Surprisingly, these variations of the PE transients in
TLS were only reported in ruby \cite{SamartsevJETP1978}, where the
characteristic excitation and depolarization times are in the range
of tens of nanoseconds, i.e. orders of magnitude longer than in
semiconductor nanostructures. So far, the PE transients in QDs were
considered in the conventional $\tau_{12}-2\tau_{12}$ picture only.

\begin{figure}[t]
\includegraphics[width=\linewidth]{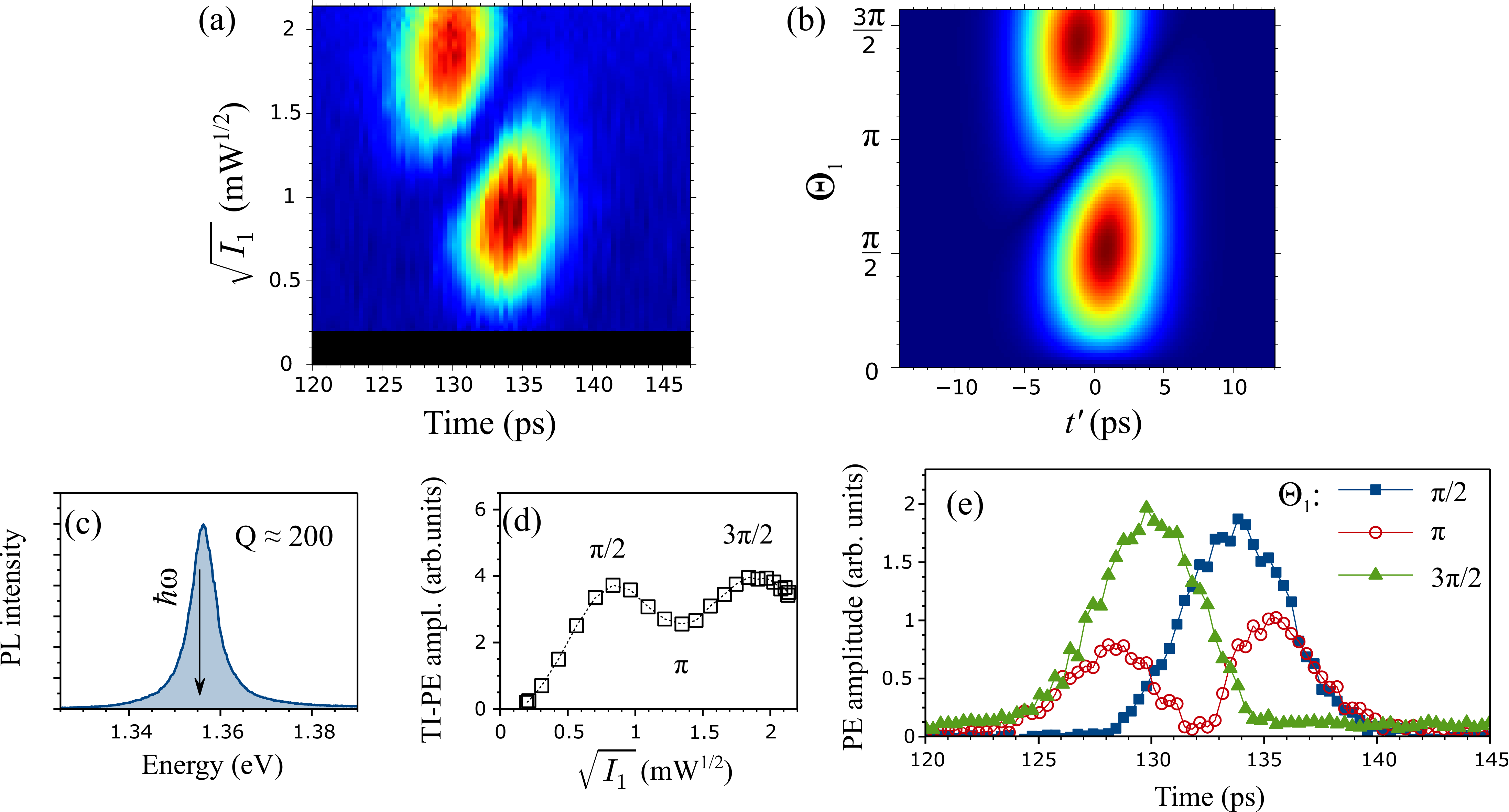}
\caption{(Color online) Experimental data and theoretical
calculations of the absolute value of PE amplitude. (a) Experimental
dependence of the PE temporal profile on the first pulse amplitude
$\sqrt{I_1}$. (b) Theoretical modeling of the data from (a) using
Eq.~(\ref{pdet}) with the following parameters: pump duration
$\tau_p = 2.9$~ps, inhomogeneity
$2\sqrt{2\ln2}\hbar\delta\approx0.65$~meV, and rephasing pulse area
$\Omega_2\tau_p=\pi$. In the colormaps, dark-blue indicates the
zero level, dark-red corresponds to maximum signal. (c) PL spectrum
of the sample with the energy position of excitation, $\hbar\omega$,
indicated by the arrow. (d) Time-integrated (TI) modulus of the PE
amplitude measured as function of the first pulse amplitude. (e) PE
transients for $\Theta_1=\pi/2$, $\pi$ and $3\pi/2$, respectively.}
\label{experimental_data}
\end{figure}

In this Letter, we demonstrate that the PE transients are strongly
influenced by the inhomogeneity of a QD ensemble. We study the
dynamics of the PE from an ensemble of (In,Ga)As/GaAs semiconductor QDs
excited by picosecond laser pulses. By varying the amplitude of the
first excitation pulse we find a complex evolution of Rabi
oscillations in the amplitude of the PE, deviating strongly from the
$\tau_{21}-2\tau_{21}$ expectation. Namely, we observe temporal
shifts of the PE pulse relative to the expected PE arrival time. We
analyze the experimental results measured on excitons using a
transparent analytical model.


The PE experiment was carried out on a sample with a
single layer of (In,Ga)As QDs which is inserted into a GaAs
$\lambda$-microcavity. The QD density is about $1.8\times10^9$
cm$^{-2}$ and the barrier contains a Si $\delta$-layer with donor
density $8\times10^9$ cm$^{-2}$, separated from the QD-layer by 10 nm.
The Bragg mirrors consist of alternating GaAs and AlAs layers with thicknesses of 68
nm and 82 nm, respectively. They have 5 and 18 such pairs in the top and
bottom mirror, respectively. With a microcavity $Q$-factor of
$\approx200$, the photon lifetime is much shorter than the laser pulse
duration of several picoseconds. Therefore, the main influence of the
microcavity is to enhance the light-matter interaction for those QDs with
optical frequency in resonance with the cavity mode. Further, the 
broad exciton photoluminescence (PL)
spectrum from the QDs with a width 
of 30~meV is convoluted with the relatively narrow photon cavity mode to result in a relatively narrow 
emission band with full-width of 6.5~meV. The energy of the cavity mode depends
on the sample spot due to the wedged cavity design. A PL spectrum
measured at the studied sample spot discussed subsequently is shown in
Fig.~\ref{experimental_data}(c).

To study the PE dynamics, a degenerate transient four-wave-mixing (FWM) experiment 
was performed in reflection geometry. The sample was placed into a helium bath cryostat 
and cooled down to a temperature of $\sim2$~K. We used a tunable mode-locked Ti:Sa laser 
as source of the optical pulses with duration of about 2.5~ps and repetition rate of about 76 MHz. 
We implemented an interferometric heterodyne detection using a reference beam to detect the absolute 
value of the FWM amplitude with temporal resolution. The excitation pulses were linearly 
cross-polarized resulting a linearly polarized PE signal 
\footnote{Accurate theoretical description requires use of a 4$\times$4-density matrix formalism 
in order to account for the linear polarizations as shown in Ref.~\cite{LangerPRL}. 
It was checked, however, that the effect of the PE shifts demonstrated here is valid for any polarization configuration.}. 
A detailed description of the experimental technique can be found elsewhere \cite{LangerPRL}.

In the experiment, the temporal profiles of the PE were measured as a function of the laser pulse amplitude. 
Figure~\ref{experimental_data} summarizes the experimental data obtained when the amplitude of the first excitation pulse is varied. 
This pulse amplitude is proportional to $\sqrt{I_1}$, where $I_1$ is the time-integrated intensity of the laser pulses. 
The delay between the centers of rephasing and excitation pulses is set to 66.7~ps. 
Clearly, footprints of Rabi oscillations are observed in Fig.~\ref{experimental_data}(a). Most strikingly, 
sizable temporal shifts of the PE maxima for $\pi/2$ and $3\pi/2$-excitation pulses (around 0.8 mW$^{1/2}$ and 1.8 mW$^{1/2}$, respectively) as well as a splitting of the PE profile around $\pi$-pulse amplitude (around 1.3 mW$^{1/2}$) become evident 
in Fig.~\ref{experimental_data}(e), where the corresponding PE transients are plotted. 
The time-integrated PE signal also manifests Rabi oscillations, 
as seen from Fig.~\ref{experimental_data}(d), but the oscillations are much less pronounced 
than in the time-resolved data. Note, that variation of the second pulse amplitude does not lead 
to any additional modification of the PE temporal profile.


In order to describe the PE from a QD ensemble we calculate the temporal evolution of a set of 
2$\times$2 density matrices interacting with two sequent laser pulses and analyze the resulting macroscopic polarization 
of the medium, $P$, given by the mean value of the dipole moment operator $\hat{d}\rho$:
\be
P = Tr(\hat{d}\rho).
\ee
\noi Here, $\rho$ is the density matrix of the TLS and $Tr$ is the trace of the matrix. 
The temporal evolution of the density matrix is described by the Lindblad equation: 
$i\hbar\dot{\rho}=[\hat{H},\rho]+\Gamma$, where $\hat{H}$ is the Hamiltonian of the system and 
$\Gamma$ describes relaxation processes phenomenologically. We consider the case, when the individual oscillator 
coherence time $T_2 \gg \tau_{12}$, 
which is justified by $T_2$ being ultimately limited by the exciton lifetime of about 0.5~ns.
Thus, the only consequence of relaxation in the system is the PE exponential decay $\sim \exp(-2\tau_{12}/T_2)$. 
Therefore, we will neglect relaxation considering $\Gamma=0$, for simplicity.

In our model, the Hamiltonian consists of two parts: $\hat{H}=\hat{H}_0+\hat{V}$, where $\hat{H}_0$ describes 
the unperturbed system and $\hat{V}=-\hat d E$ gives the interaction with light. We assume rectangular shaped pulses, each with
with duration $\tau_p$. During the optical excitation, we calculate the ensemble dephasing due to the detuning of every individual 
oscillator with frequency $\omega_0$ from the light frequency $\omega$: $\Delta=\omega-\omega_0$. 
In the rotating-wave approximation, this defines the optical nutation of the TLS on the Bloch sphere 
with a generalized Rabi frequency $\tilde{\Omega}=\sqrt{\Omega^2+\Delta^2}$, where $\Omega=2|E_0 d|/\hbar$ is proportional to the 
optical field amplitude $E_0$. Accordingly, the analysis of the coherent optical response of the system includes the following steps: 
first, we calculate the interaction of the unperturbed system with the first excitation pulse ($\Omega=\Omega_1$) 
acting during time $\tau_p$, 
then the system evolves freely in absence of excitation ($\Omega=0$) during the time interval $\tau_{12}$, 
followed by the excitation with the second rephasing pulse ($\Omega=\Omega_2$) during time $\tau_p$. 
Finally, we analyze the macroscopic polarization of the QDs that is available for PE pulse formation 
by integrating the individual responses over the inhomogeneous ensemble of TLS, defined by a Gaussian distribution 
with standard deviation $\delta$. Within these approximations, it is possible to perform the analysis analytically 
and calculate the following expression for the modulus of the detected excitation-dependent part of the PE amplitude 
(See Appendix A containing details of the theoretical approach.):

\begin{align}
&|P_{PE}(t')| \sim \bigg|\displaystyle\int_{-\infty}^{\infty} \exp\bigg(-\frac{\Delta^2}{2\delta^2}\bigg) \frac{\Omega_1\Omega_2^2}{\tilde{\Omega}_1\tilde{\Omega}_2^2}\sin^2\frac{\tilde{\Omega}_2\tau_p}{2} \nonumber \\
&\times \bigg[\frac{2\Delta}{\tilde{\Omega}_1} \sin^2\frac{\tilde{\Omega}_1 \tau_p}{2} \sin\Delta t' + \sin\tilde{\Omega}_1 \tau_p\ \cos\Delta t' \bigg] d\Delta\bigg|.
\label{pdet}
\end{align}

\noi Here $t'=t-2(\tau_{12}+\tau_p)$, and $\tilde{\Omega}_{1,2}$ are the generalized Rabi frequencies of the first and the second pulse, 
respectively. In this notation, delay between the exciting pulse centers used in experiment is $\tau_{12}+\tau_p$ and expected PE arrival time according to the standard treatment is $t'=0$. This equation is similar to the one obtained for spin echoes measured in highly inhomogeneous fields \cite{Bloom1955,Tsifrinovich1998}. 
It can be easily seen, that variation of the second, rephasing pulse amplitude and, hence, $\Omega_2$ 
does not affect the symmetry of the PE profile. Therefore, we will consider only the effect coming from variation of the first, 
excitation pulse amplitude.

\begin{figure}[t]
\includegraphics[width=0.8\linewidth]{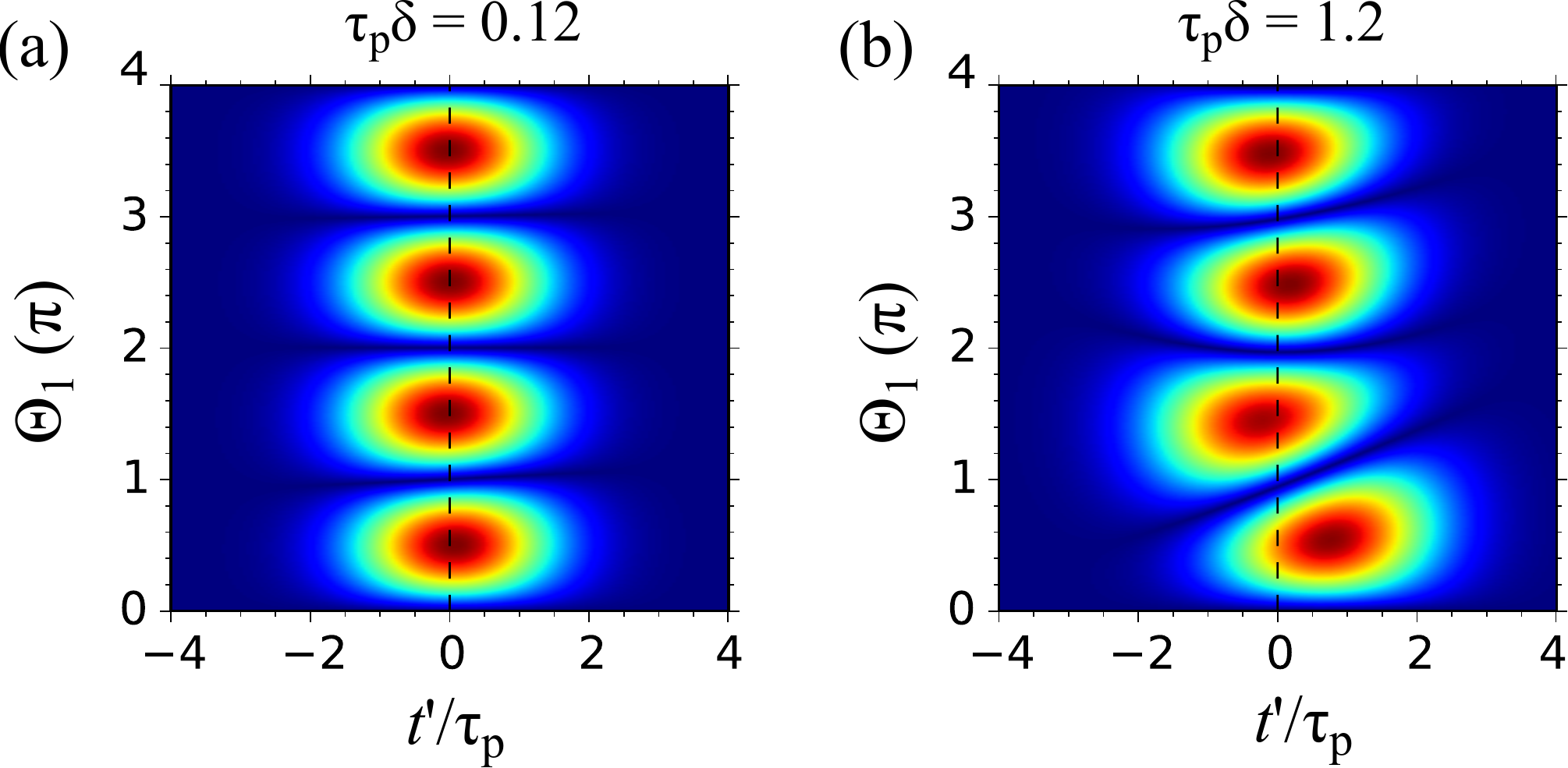}
\caption{(Color online) Theoretical calculation of the PE temporal profile as function of the excitation pulse area $\Theta_1$ 
using Eq.~(\ref{pdet}) for two different ensemble with different inhomogeneity $\delta$:
(a)~$\tau_p\delta=0.12$; and (b)~$\tau_p\delta=1.2$. The time scale is normalized by the pulse duration. 
Dark-blue corresponds to zero signal level, dark-red gives the maximum value.}
\label{simulations}
\end{figure}

Figure \ref{simulations} shows results of theoretical calculations performed using Eq.~(\ref{pdet}) for two ensembles 
with different widths of the inhomogeneity: $\tau_p\delta=0.12$ and 1.2 \footnote{In the calculations, dimensionless units are used 
for convenience 
for the frequency and the time scales. In these units, the pulse spectral width is equal to 1.}. 
For an ensemble broadening much narrower than the pulse spectrum, $\tau_p\delta=0.12$, clean Rabi oscillations are seen in the PE 
temporal profile as function of the excitation pulse area $\Theta_1=\Omega_1\tau_p$, shown in Fig.~\ref{simulations}(a). 
In particular the photon echo arrives as expected at twice the separation between rephasing and excitation pulse. 
This case is equivalent to an ideal Hahn echo, where the echo signal is generated
at $2 \tau_{12}$. 
When, however, the inhomogeneous ensemble width becomes comparable with the pulse spectral width ($\tau_p\delta=1.2$), 
strong modifications of the temporal PE profile appear, as can be seen in Fig.~\ref{simulations}(b). 
The key feature of these modifications is the effective temporal shift of the PE amplitude that is particularly prominent  
for the first two maxima of the Rabi oscillations when increasing the excitation power: 
for a $\pi/2$-pulse the PE maximum is retarded relative to the expected PE arrival time, while for a
$3\pi/2$-pulse the PE is advanced. 
These shifts do not depend on the pulses delay time $\tau_{12}$ in the limit of negligible damping 
and decrease with further increasing $\Theta_1$.

\begin{figure}[t]
\includegraphics[width=0.7\linewidth]{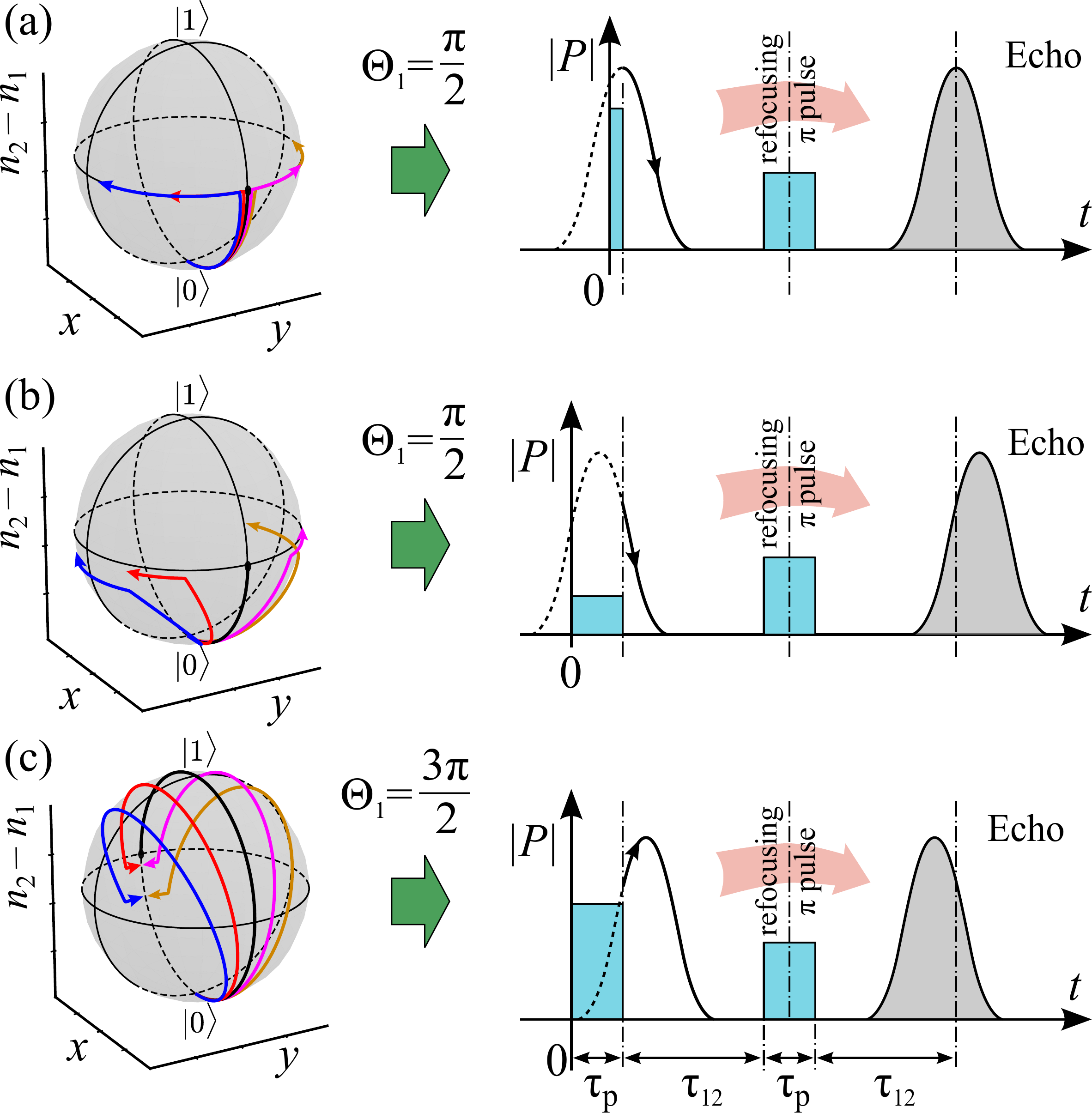}
\caption{(Color online) Explanation of the PE timing modifications using the Bloch sphere representation for different excitation 
pulse configurations: (a) ultrashort $\pi/2$-pulse: case of Hahn echo. (b) $\pi/2$-pulse: retarded echo. (c) $3\pi/2$-pulse: 
advanced echo. The colored lines on the Bloch spheres (left panels) denote different trajectories of the Bloch vectors of 
individual oscillators with different detuning from the laser frequency in directions indicated by the arrows. The right panels show 
the temporal evolution of the modulus of the macroscopic polarization, $|P|$, after excitation pulse action. The dashed lines denote 
extrapolation of $|P|$ to $t<\tau_p$.}
\label{drift_explanation}
\end{figure}

The temporal shifts of the PE can be qualitatively explained using the Bloch sphere representation, as outlined in 
Fig.~\ref{drift_explanation}. In Fig.~\ref{drift_explanation}(a), the ideal case of a Hahn echo is given as a reference, 
where excitation with a negligibly short excitation $\pi/2$-pulse results in an unshifted echo signal. 
When the excitation pulse has a finite duration, as shown in Fig.~\ref{drift_explanation}(b), oscillators experience dephasing 
during pulse action, so that the excitation distributes the oscillators along the equator of the Bloch sphere 
followed by their precession in the equatorial plane. This precession extrapolated to negative times results in focusing 
all the Bloch vectors on the same meridian, which corresponds to a situation where the maximum macroscopic polarization is 
shifted to $t<\tau_p$. 
In case of excitation by a $3\pi/2$-pulse as shown in Fig.~\ref{drift_explanation}(c), after pulse action all 
oscillators proceed along the equator towards each other, which leads to a shift of the macroscopic polarization maximum to $t>\tau_p$. 
Since the rephasing pulse, which is not necessarily a $\pi$-pulse, 
acts essentially as a refocusing pulse, which in effect reverses the temporally coherent evolution of the ensemble oscillators, 
the dynamics of the macroscopic dipole moment around the expected echo arrival time, $2(\tau_{12}+\tau_p)$, is similar to a 
mirrored copy of its dynamics around the time $t=\tau_p$, as shown by the right panels of Fig.~\ref{drift_explanation}. 
Thus, the PE pulse is retarded for the a $\pi/2$-excitation pulse and advanced for a $3\pi/2$-pulse, when the ensemble dephasing 
during pulse action is sufficiently strong. This can be achieved either by high inhomogeneity in the excited ensemble 
and/or long pulse duration.

The analysis of the experimental data in Fig.~\ref{experimental_data}(a) was performed using the theoretical modeling of the PE effect 
at varying excitation pulse amplitudes after Eq.~(\ref{pdet}). The adjustable parameters are $\delta$ and $\tau_p$. 
The theoretical fit depicted in Fig.~\ref{experimental_data}(b) gives a pulse duration $\tau_p\approx2.9$~ps and an 
inhomogeneous ensemble full-width at half-maximum $2\sqrt{2\ln2}\hbar\delta\approx0.65$~meV. This broadening is much smaller 
than the PL linewidth of the QD structure (6.5~meV) indicating that a much narrower sub-ensemble contributes to the PE compared to  
the ensemble contributing to the PL emission. The origin of this discrepancy will be subject of further studies. A possible reason might be the interaction with 
the microcavity. The evident Rabi oscillations pattern in the PE sequence tells that QD structures under resonant excitation 
can be well described by TLS.


In conclusion, the importance of the oscillator detuning from the optical frequency in forming the macroscopic coherent response 
in semiconductor QDs was addressed in literature already before \cite{LangbeinPRL2005,RamsayPRB}. However, 
no consequence of the timing of the associated PE had been reported. 
The essential consequence of our study is that even significantly inhomogeneous semiconductor systems still demonstrate 
a comprehensible coherent behavior, but great care needs to be exercised to consider the consequences of the inhomogeneity in detail. 
When doing so, such systems are very well suited candidates for the coherent manipulations in coherent optoelectronics. 
In particular, the observed temporal shifts of the PE amplitude need to be taken into account 
when a precise timing of the optical signals matters. Our findings have strong impact on interpretation of two-dimensional Fourier transform images, which are currently widely used because of their intuitive visualization of underlying physics \cite{Moody2013, Cundiff2009}. The complex-valued envelope of the PE-transients from strongly inhomogeneous systems results in a changeover of the corresponding 2D image from a resonance profile in case of a Hahn echo to a dispersive profile (see Appendix B).

\section*{\label{sec:level1}Acknowledgments}

Authors thank Gleb Kozlov for fruitful discussions. We acknowledge the financial support by the Deutsche Forschungsgemeinschaft 
and the Russian Foundation of Basic Research in the frame of the ICRC TRR 160 (RFBR project No. 15-52-12016 NNIO\_a). The project ‘SPANGL4Q’ acknowledges financial support from the Future and Emerging Technologies (FET) programme within the Seventh Framework Programme for Research of the European Commission, under FET-Open grant no. FP7-284743. S.V.P. thanks the Russian Foundation of Basic Research for partial financial support (contract no. 14-02-31735 mol\_a). S.V.P. and I.A.Yu. acknowledge partial financial support from the Russian Ministry of Science and Education (contract no. 11.G34.31.0067) and SPbU (Grant No. 11.38.213.2014). 
M.B. acknowledges support from the Russian Ministry of Science and Education (contract number 14.Z50.31.0021).

\section*{Appendix A: Theoretical model}

We consider optical excitation of an inhomogeneous ensemble of two-level systems by two short laser pulses with central frequency $\omega$ close to the average frequency of the ensemble $\bar{\omega}_{0}$.
For each individual two-level system the incident electromagnetic field induces optical transitions creating a coherent superposition of states.  Let $|0\rangle$ denote the ground state and $|1\rangle$ denote the dipole active, excited state. 

The temporal evolution of the density matrix is described by the Lindblad equation:
\begin{equation}
\label{eq:eq1}
\dot{\rho}=-\frac{i}{\hbar}[\hat{H},\rho]+\Gamma.
\end{equation}
Here $\hat{H}=\hat{H}_0+\hat{V}$, where $\hat{H}_0$ is the
Hamiltonian of the unperturbed system. $\Gamma$
describes relaxation processes phenomenologically. 
$\hat{V}$ is the perturbation operator describing
the interaction with the electromagnetic wave of a pulse: $V_{10} = V_{01}^* = -d E(t) \exp(-i \omega t)$,  $d$ is the dipole moment of the optical transition to state $|1\rangle$, $E(t)$ is the envelope of the electric field of the pulse. Here the rotating wave approximation is used. For simplicity, we consider pulses with rectangular shape of amplitude $E_0$ and assume that the pulse duration, $\tau_p$, is significantly shorter than all relaxation processes.

At the end of a short pulse, the density matrix elements are \cite{LangerPRL}:
\bea
\rho_{00}(\tau_p)&=&\rho_{00}(0)-\frac{\Im{\big(f\rho_{01}(0)\big)}}{\tilde{\Omega}}\sin\tilde{\Omega} \tau_p \nonumber \\
&+& (1-\cos\tilde{\Omega} \tau_p)\frac{\left[\Delta\Re\big(f\rho_{01}(0)\big)+\big(\rho_{11}(0)-\rho_{00}(0)\big)|f|^2/2\right]}{\tilde{\Omega}^2},\nonumber\\
\rho_{11}(\tau_p)&=&\rho_{11}(0)+\frac{\Im{\big(f\rho_{01}(0)\big)}}{\tilde{\Omega}}\sin\tilde{\Omega} \tau_p \nonumber \\
&-& (1-\cos\tilde{\Omega} \tau_p)\frac{\left[\Delta\Re\big(f\rho_{01}(0)\big)+\big(\rho_{11}(0)-\rho_{00}(0)\big)|f|^2/2\right]}{\tilde{\Omega}^2},\nonumber\\
\rho_{01}(\tau_p)&=&
\exp(i\omega \tau_p)\bigg[\rho_{01}(0)\Big[\cos\tilde{\Omega} \tau_p - i\frac{\Delta}{\tilde{\Omega}} \sin\tilde{\Omega} \tau_p\Big] + \frac{f^*\Re\big(f\rho_{01}(0)\big)}{\tilde{\Omega}^2}(1-\cos\tilde{\Omega} \tau_p)\nonumber\\
&-&{f^*\big(\rho_{11}(0)-\rho_{00}(0)\big)\over 2\tilde{\Omega}^2}\left[\Delta(1-\cos\tilde{\Omega} \tau_p)+i\tilde{\Omega} \sin\tilde{\Omega} \tau_p\right]\bigg].
\label{eq:eq5a}
\eea

Here, $\Delta=\omega-\omega_0$ is the detuning between the optical frequency
$\omega$ and the resonance frequency of the individual two-level system $\omega_0$. 
$f$ is proportional to the smooth envelope of the pulse:
\[
f(t) = -\frac{2d E(t)}{\hbar}.
\]
In the rectangular pulse approximation, $f$ is constant during the pulse action, thus, $\tilde{\Omega}=\sqrt{|f|^2+\Delta^2}\equiv \sqrt{\Omega^2+\Delta^2}$, where $\Omega$ is the Rabi frequency.

Let us consider now the interaction of the two-level system with two pulses separated by the time, $\tau_{12}$.
We assume that before excitation with the first pulse all two-level systems are in the ground state $|0\rangle$ 
and, therefore, only $\rho_{00}(0) \ne 0$.
The density matrix elements after this first pulse are \cite{LangerPRL}:
\bea
\rho_{00}(\tau_p)&=&\rho_{00}(0) 
-(1-\cos\tilde{\Omega}_1 \tau_p)\frac{\rho_{00}(0)|f_1|^2}{2\tilde{\Omega}_1^2},\nonumber\\
\rho_{11}(\tau_p)&=&
(1-\cos\tilde{\Omega}_1 \tau_p)\frac{\rho_{00}(0)|f_1|^2}{2\tilde{\Omega}_1^2},\nonumber\\
\rho_{01}(\tau_p)&=&
\exp(i\omega \tau_p)
{f_1^*\rho_{00}(0)\over 2\tilde{\Omega}_1^2}\left[\Delta(1-\cos\tilde{\Omega}_1 \tau_p)+i\tilde{\Omega}_1 \sin\tilde{\Omega}_1 \tau_p\right].
\label{eq:eq5a}
\eea
Here the index $1$ indicates the first pulse.
After the pulse, the elements of the density matrix relax with different characteristic times. 
We assume that $\rho_{01}$ describing the optical coherence relaxes with the dephasing time $T_{2}$: 
$\rho_{01} = \rho_{01}(\tau_{p}) \exp\big[i\omega_0 (t-\tau_p)\big]\exp\big[-(t-\tau_p)/T_{2}\big]$.

After the action of the second pulse we need only the following term in $\rho_{01}$ for calculation of the photon echo: 
\bea 
\rho_{01}(\tau_{12} +2\tau_p)
&\propto& \frac{(f_{2}^*)^2
\rho_{10}(\tau_{12}+\tau_p)\sin^2(\tilde{\Omega}_{2}\tau_p/2)}{\tilde{\Omega}_{2}^2}\exp(i\omega \tau_p).
\eea

Finally, after two pulses:
\bea
\rho_{01}(t)&\propto&\frac{f_1^*(f_{2}^*)^2 \exp\big[i\omega_0 (t-2(\tau_{12}+\tau_p))\big]\exp(-(t-2\tau_p)/T_{2})}{2\tilde{\Omega}_1^2\tilde{\Omega}_{2}^2} \nonumber \\
&\times& \left[\Delta(1-\cos\tilde{\Omega}_1 \tau_p)-i\tilde{\Omega}_1 \sin\tilde{\Omega}_1 \tau_p\right]
\sin^2(\tilde{\Omega}_{2}\tau_p/2).
\eea 

The measured photon echo signal from the ensemble of two-level systems is proportional 
to the  sum over all $\omega_0$: $\big|\sum_{\omega_0} \rho_{01}(\omega_0)F(\omega_0)\big|$, where $F(\omega_0)$ is
the distribution function of the ensemble.
If the distribution function is Gaussian with a central frequency $\bar{\omega}_0=\omega$ and a
dispersion $\delta$, this gives the following form of the echo signal:
 
\bea
\big|&P_{PE}(t)&\big|\propto \left|\int_{-\infty}^{\infty}
\frac{\Omega_1\Omega_{2}^2\exp(-(t-2\tau_p)/T_{2})\sin^2(\tilde{\Omega}_{2}\tau_p/2)}{\tilde{\Omega}_1^2\tilde{\Omega}_{2}^2}\exp(-\Delta^2/2\delta^2) \right. \nonumber \\
&\times& \left. \left[\Delta(1-\cos\tilde{\Omega}_1 \tau_p)\sin\big[\Delta(t-2(\tau_{12}+\tau_p))\big]
+\tilde{\Omega}_1 \sin\tilde{\Omega}_1 \tau_p \cos\big[\Delta(t-2(\tau_{12}+\tau_p))\big]\right]d\Delta \right|. \nonumber \\
\label{P_PE_abs}
\eea

\section*{Appendix B: Photon Echo in 2D Fourier Spectroscopy}

In order to calculate the two-dimensional Fourier spectrum (2DFS) of a photon echo (PE), we integrate the PE field over the two temporal varaibles that determine the delay between the laser pulses, $t_1=\tau_{12}$, and the laboratory time, $t_3=t-\tau_{12}$. To that end we rewrite Eq.(\ref{P_PE_abs}):

\bea
P_{PE}(t_1,t_3)&\propto& \int_{-\infty}^{\infty}
\frac{\Omega_1\Omega_{2}^2\exp(-(t_3-t_1)/T_{2})\sin^2(\tilde{\Omega}_{2}\tau_p/2)}{\tilde{\Omega}_1^2\tilde{\Omega}_{2}^2}\exp(-\Delta^2/2\delta^2) \nonumber \\
&\times& \left[\Delta(1-\cos\tilde{\Omega}_1 \tau_p)\sin\big[\Delta(t_3-t_1-2\tau_p)\big]
+\tilde{\Omega}_1 \sin\tilde{\Omega}_1 \tau_p \cos\big[\Delta(t_3-t_1-2\tau_p)\big]\right]d\Delta \nonumber. \\
\label{P_PE}
\eea

\noi Then, the two-dimensional Fourier transform is taken according to the following expression:

\bea
S(\omega_1,\omega_3)\propto\int\int_{-\infty}^{\infty}P_{PE}(t_1,t_3)\exp(i\omega t_3-i\omega t_1)\exp(i\omega_1 t_1+i\omega_3 t_3)dt_1dt_3.
\label{2DFTS}
\eea

\begin{figure}[h]
\includegraphics[width=0.7\linewidth]{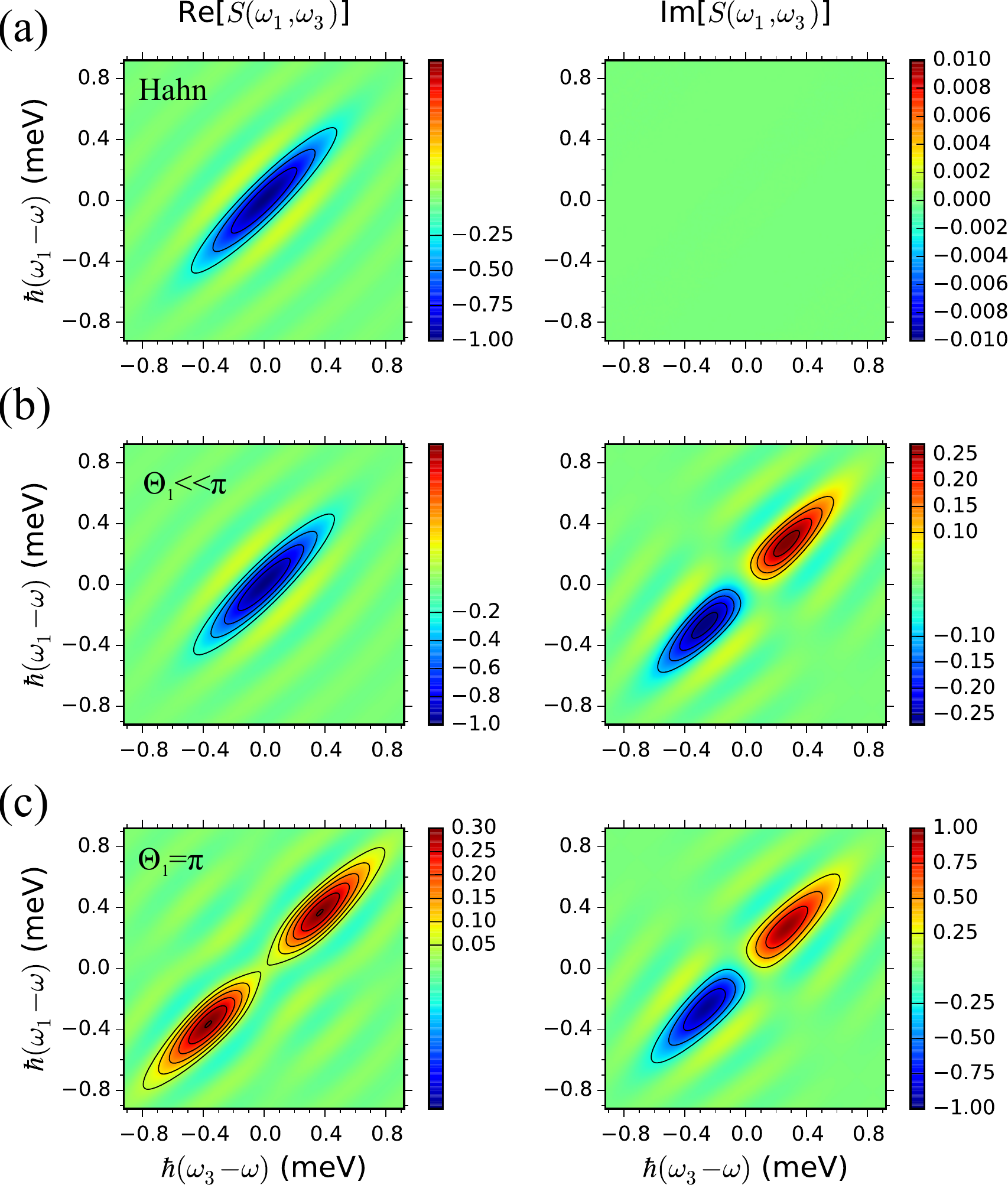}
\caption{(Color online) Results of 2DFS simulations using Eq.(\ref{2DFTS}) for three different cases: (a) Hahn-echo; (b) linear excitation regime with weak intensity of the first pulse, $\Theta_1=0.01\pi$; (c) excitation in the non-linear regime with the first pulse being a $\pi$-pulse. The 2DFS spectra are normalized to unity. Left and right panels denote the real and imaginary parts of the $S(\omega_1,\omega_3)$ spectra, respectively, plotted in the same scaling. Other parameters in the calculations are: $2\sqrt{2\ln2}\hbar\delta=0.65$~meV, $\tau_p=2.5$~ps.}
\label{som_fig}
\end{figure}

Figure~\ref{som_fig} demonstrates the results of corresponding 2DFS calculation after Eq.(\ref{2DFTS}) for TLS ensemble excitation with different areas, $\Theta_1$ of the first pulse. The 2DFS spectra are normalized to unity. The first case shown in Fig.~\ref{som_fig}(a) accords with a traditional Hahn echo, for which no temporal shift of the PE is expected. Here, the 2DFS has only a real part. 
Note that the weak fringes in the plot are an artifact due to taking the Fourier transform over finite intervals.
In case of weak excitation (linear regime) of a strongly inhomogeneous system similar to the one studied experimentally, the temporal shift of the PE gives rise to a significant imaginary part of the 2DFS with a dispersive shape and a sign reversal along the diagonal, as shown in Fig.~\ref{som_fig}(b). 
When increasing the power of the first excitation pulse further to a $\pi$-pulse, see Fig.~\ref{som_fig}(c), also the real part of the 2DFS undergoes strong changes: its sign becomes reversed compared to the weak power regime and it shows a double peak structure, while the imaginary part of the spectrum is now the dominant contribution.

\end{document}